\documentclass[twocolumn,twoside,slac_two]{revtex4}
\usepackage{graphicx}
\usepackage{fancyhdr}
\def\met{$E_{\rm T}^{\rm miss}$~}
\def\pt{$p_{\rm T}$~}
\def\meff{$M_{\rm eff}$~}

\def\mlledge{m_{\ell \ell}^{\rm edge}}

\def\mllqedge{m_{\ell \ell q}^{\rm edge}}
\def\mllqthr{m_{\ell \ell q}^{\rm thr}}
\def\mlqlowedge{m_{\ell q (\rm{low})}^{\rm edge}}
\def\mlqhighedge{m_{\ell q (\rm{high})}^{\rm edge}}
\def\chione{\widetilde{\chi}^{0}_{1}}
\def\chitwo{\widetilde{\chi}^{0}_{2}}
\def\squark{\widetilde {q}_{L}}
\def\slepton{\widetilde {l}_{R}}

\begin{document}

\title{Search for Supersymmetry Signatures at the LHC}

%

\author{N. \"Ozt\"urk, for the ATLAS and CMS Collaborations}
\affiliation{Department of Physics, University of Texas at Arlington, Arlington, TX, 76019, USA}
\begin{abstract}
Supersymmetry (SUSY) is one of the most interesting and comprehensively studied models for new 
physics beyond the Standard Model. If SUSY exists in nature the Large Hadron Collider will provide 
excellent opportunities to search for SUSY. SUSY discovery strategies of the ATLAS and CMS experiments 
are presented with a focus on early data. SUSY mass measurement techniques and determination 
of SUSY model parameters are also demonstrated.
\end{abstract}

\maketitle

\thispagestyle{fancy}


\section{Introduction}
Supersymmetry is a theoretically attractive model for extention of the Standard Model 
(SM)~\cite{SUSY}. It proposes that all the SM particles have SUSY partners with a spin 
difference of $\pm 1/2$. Among its motivations; SUSY solves the fine-tuning problem 
(Higgs mass stabilization against loop corrections), SUSY modifies running of the 
SM gauge couplings just enough to give the grand unification at a single scale 
and SUSY also offers a candidate for dark matter through R-parity conservation. 
Since SUSY partners of the SM particles have not been observed at the same 
mass scale SUSY must be a broken symmetry at low energy. The Minimal Supersymmetric 
Standard Model (MSSM) brings more than a hundred free parameters into the theory, 
thus searching for SUSY is a very challenging task. 

One of the main objectives for building the Large Hadron Collider (LHC) is to discover SUSY. 
The ATLAS (A Toroidal LHC ApparatuS) and CMS (Compact Muon Solenoid) are two general 
purpose detectors at the LHC designed to search for SUSY particles (and for the Higgs 
boson). Both experiments have successfully seen the first beam events in September 
2008 and have been commissioning with the cosmic rays since then. The LHC is scheduled 
to start up in November 2009.

The ATLAS and CMS experiments put enormous efforts recently to get ready for the analysis
of the first LHC data. ATLAS has written a collection of notes published as a CERN 
book~\cite{Aad:2009wy} based on the results from the Computer System Commissioning (CSC)
exercise. CMS has written a Technical Design Report II~\cite{CMS:TDR}. Both have dedicated
chapters to SUSY covering a broad spectrum in SUSY analyses with 14 TeV center of mass
energy. Post-CSC studies in ATLAS~\cite{ATLASnotebjet, ATLASnoteTilesMethod, ATLASnotettbckg10TeV, 
ATLASnoteInclusive10TeV,ATLASnoteSS10TeV} and post-TDR studies in 
CMS~\cite{CMSnoteMultijet, CMSnoteDilepton10TeV, CMSnoteDiphoton, CMSnoteGluino, CMSnotebackgZnunu} 
have followed for both 14 TeV and 10 TeV center of mass energies resulting in a variety of public notes.
This review article is based on these recent studies. After a brief mention of the SUSY 
search strategies at the LHC, examples of the data-driven methods will be given for 
estimating the SM backgrounds to SUSY searches. Then searches with the inclusive 
missing transverse energy ($E_{\rm T}^{\rm miss}$) signatures and an exclusive search without explicity 
use of \met will be summarized. Examples from special SUSY signatures of the non-standard SUSY models 
will be presented together with the discovery reach of some SUSY models. The article will conclude  
after a discussion on the SUSY mass measurements and determination of SUSY model 
parameters upon a possible SUSY discovery.

\section{SUSY Search Strategies at the LHC}
The SUSY breaking mechanism determines the phenomenology and search strategy in the 
collider experiments. Many possible mechanisms have been proposed; mSUGRA~\cite{mSUGRA}, 
GMSB~\cite{GMSB}, AMSB~\cite{AMSB}, SO(10) SUSY GUTs~\cite{SO10}, Split-SUSY~\cite{SplitSUSY} 
and others. The search strategies can be discussed in two classes of models:
\begin{enumerate}
\item Standard SUSY models with R-parity conservation. In this class SUSY is searched for 
as an evidence of excess above the SM background in the following channels; 
0, 1, 2 leptons+$\geq$ 2, 3, 4 jets+$E_{\rm T}^{\rm miss}$, $\gamma$+jets+$E_{\rm T}^{\rm miss}$, 
$\tau$ or $b$-jets+$E_{\rm T}^{\rm miss}$, multi-leptons, multi-jets without \met and others.
\item Non-standard SUSY models with special signatures; displaced vertices, stopped gluinos 
in the calorimeter, non-pointing photons, long-lived stable massive particles and others.
\end{enumerate}

Understanding the SM backgrounds play an essential role in SUSY searches. Main sources 
of the SM backgrounds come from the following processes; $t \bar t$+jets, $W$+jets, $Z$+jets, 
QCD jets and diboson ($ZZ$, $WW$, $WZ$) production. The data-driven estimation of the 
SM backgrounds, will be an emphasis of early analyses. Understanding the fake \met and tails 
in the \met distributions as well as the reconstruction of the non-standard signatures 
will be a critical task in the search for SUSY. 

\section{Estimation of the Standard Model Backgrounds to SUSY Searches}
Several Monte Carlo-driven and data-driven estimation techniques have been developed 
to estimate the SM backgrounds. Some of the data-driven estimation techniques are listed as: 

\begin{itemize}
\item $Z \rightarrow \nu \bar \nu$ background in 0-lepton mode~\cite{CMSnotebackgZnunu}.
\item QCD background in 0-lepton mode SUSY search using the jet smearing method~\cite{Aad:2009wy}.
\item QCD background in 1-lepton mode SUSY search using the lepton isolation method~\cite{Aad:2009wy}.
\item $t \bar t$ background in 0-lepton mode using the replacement technique~\cite{ATLASnotettbckg10TeV}.
\item $t \bar t$ background in 1-lepton mode SUSY search using the Topbox method~\cite{Aad:2009wy}.
\item Combined background in 1-lepton mode SUSY search using $m_T$ method~\cite{Aad:2009wy}
(further development with combined fit) and Tiles method~\cite{ATLASnoteTilesMethod}.
\item Others.
\end{itemize}

As an example two of the data-driven approaches are summarized below. 
\subsection{Data-driven estimation of the QCD background in the hadronic channel}
The biggest background estimation challenge for SUSY searches in the jets+\met channels 
will be estimation of the QCD jet background. This background involves two sources of $E_{\rm T}^{\rm miss}$:
\begin{enumerate}
\item Fake $E_{\rm T}^{\rm miss}$; non-Gaussian tails to the detector jet response function arising from 
the dead material, jet punch-through, pile-up and other effects. It can be suppressed by 
applying the jet-\met azimuthal angle cuts, calorimeter and tracking cuts and cosmic 
background rejection cuts.
\item Real $E_{\rm T}^{\rm miss}$; resulting from the non-interacting particles such as neutrinos 
or the Lighest Supersymmetric Particle (LSP).  
\end{enumerate}

Two approaches are taken to estimate the remaining QCD jet backgrounds; Monte Carlo-driven 
estimates and data-driven estimates. The Monte Carlo based estimates 
are subject to large systematics effects arising from the proton parton distributions, 
underlying event uncertainties, jet energy scale uncertainty and an uncertainty in 
modeling the QCD jet physics with Monte Carlo generators (PYTHIA versus ALPGEN). 
There will also be an uncertainty from the luminosity measurement. In addition, the detector 
simulation uncertainties (imperfect description of the 
response of the detector to jets) and the statistical uncertainties (large QCD cross 
section rendering production of full simulation samples unfeasible) will contribute. 
As for the data-driven estimates they will be crucial in early phase of data taking as 
they are less prone to input systematics. 

A method from an ATLAS study~\cite{Aad:2009wy} is described here for estimating the QCD jet background 
in the 0-lepton mode SUSY search. The method makes use of smearing the jet transverse momenta in low 
\met QCD jet data with a data measured jet response function R (ratio of the measured 
jet \pt to the true jet $p_{\rm T}$). The Gaussian part of the jet response function is measured with 
the balance of the $\gamma$+jet events. The non-Gaussian part of the jet response function is 
measured from the 'Mercedes' type of configuration (\met is parallel or anti-parallel to the 
\pt of one of the three jets selected in the event). The jet response function is then used to 
smear the jet \pt in multijet events with low \met in order to estimate the \met distribution of 
QCD multijet events. 
\begin{figure}[ht]
\centering
\includegraphics[width=70mm]{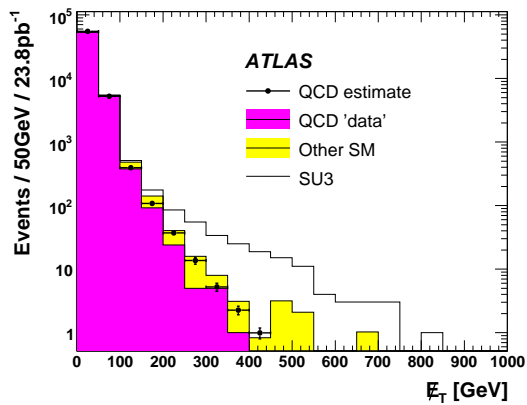}
\caption{The \met distribution of signal and background for an integrated luminosity
of 23.8~pb$^{-1}$ at 14 TeV. The QCD background is estimated from the jet smearing method
in 0-lepton SUSY search as explained in the text. The signal point is the ATLAS mSUGRA
benchmark point SU3.}
\label{jetSmearing}
\end{figure}
Figure~\ref{jetSmearing} shows the \met distribution of the full simulation data 
and the data-driven estimated for an integrated luminosity of 23.8~pb$^{-1}$ at 14 TeV. The \met 
distribution of the SUSY signal (ATLAS mSUGRA benchmark point, SU3, with parameters $m_0 =100$ 
GeV, $m _{1/2}=300$ GeV, $A_0=-300$, ${\rm tan}(\beta)=6$, ${\rm sgn}(\mu)=+$) and the 
summed background from $Z \rightarrow \nu \bar{\nu}$+jets, $W \rightarrow l \nu$+jets, 
$t \bar t$+jets are also shown. A good agreement is seen between the estimated and the full 
simulation data. Event counts yield $2.36\pm 0.09(\rm stat)\pm 1.44(\rm syst)$ for the estimated 
versus 1 for the observed. The uncertainty on the estimate is then $~$60$\%$ for 23.8~pb$^{-1}$. 
The same uncertainty is assumed for 1~fb$^{-1}$ of data.  Given the difficulty of obtaining 
accurate estimates of the QCD background, results from a number of independent techniques 
should be compared to obtain a robust QCD background estimate.
\subsection{Data-driven background estimates for SUSY diphoton search}
The GMSB model with two high \pt photons and large \met in the final state is among the 
promising models for early SUSY searches. A method from a CMS study~\cite{CMSnoteDiphoton} is described 
here to predict the \met distribution in a diphoton sample from the SM processes. Seeing 
an excess of events at high \met would indicate a signal for new physics. The physics background 
is small and comes from the $Z \gamma \gamma \rightarrow \nu \nu \gamma 
\gamma$ and $W \gamma \gamma \rightarrow l \nu \gamma \gamma$ SM processes. The instrumental 
background comes from three major sources:
	
\begin{enumerate}
\item QCD background; results from $\gamma$-jet misidentification in QCD events with no real \met 
such as photon plus jets and multijet production.
\item Electroweak background; results from $\gamma$-$e$ misidentification in events with real 
\met from $W \gamma$ and $W$+jet production. 
\item Non-beam background; results from high energy muons from cosmic rays or beam-halo. 
\end{enumerate}

\begin{figure}[ht]
\centering
\includegraphics[width=80mm]{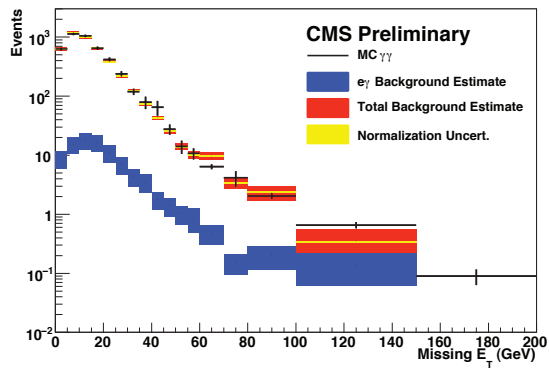}
\caption{The \met distribution of the Monte Carlo data and the data-driven estimated 
from a background closure test in absence of SUSY signal. An integrated luminosity 
of 100~pb$^{-1}$ at 10 TeV is considered.} 
\label{CMSdiphoton}
\end{figure}

It has been demonstrated that the QCD and electroweak backgrounds are under control, 
and the techniques explored will effectively eliminate the non-beam background. A CMS benchmark 
point, GM1C, is used in this study with parameters $\Lambda$=100~TeV, $M_m$=200~TeV, $N_5$=1, 
$C_G$=1, ${\rm tan}(\beta)=15$, ${\rm sgn}(\mu)=+$. Figure~\ref{CMSdiphoton} shows the \met 
distribution from the background closure test (using $Z \rightarrow e^{+}e^{-}$ events to 
describe the QCD background) in absence of SUSY signal for an integrated luminosity of 
100~pb$^{-1}$ at 10 TeV. As for comparison with the Monte Carlo truth, event counts yield 
($2.78\pm 0.24$) for the Monte Carlo $\gamma \gamma$ data and ($2.69\pm 0.66$) for the estimated 
background in absence of SUSY signal, and ($17.5\pm 0.26$) and ($2.99\pm 0.68$) respectively in 
presence of SUSY signal. A good agreement is seen between the data-driven estimated and the 
predicted background. The data-driven strategy demonstrated in this study can be used at the 
start-up of the LHC.
\section{SUSY Searches with Inclusive \met Signatures}
SUSY phenomenology varies significantly through the SUSY parameter space, therefore SUSY is 
searched inclusively based on very general signatures. It is expected that gluinos and squarks 
(heaviest SUSY particles) will be produced in the initial proton-proton interaction at the 
LHC and then their cascade decays will result in events with high \pt jets and leptons in the 
final state. In addition, if the R-parity is conserved the LSP would be stable and undetected 
leading to events with large $E_{\rm T}^{\rm miss}$. Thus typical inclusive SUSY searches cover 
a broad range of search modes depending on the presence of leptons (electron/muon), 
$\tau$ lepton or $b$ quarks in the final state; the 0-lepton mode, 1-lepton mode, 2-lepton mode, 
3-lepton mode, tau-mode and $b$-jet mode and others. Examples of these search modes are given 
in the following. A review of the SUSY searches with inclusive \met signatures from an ATLAS study 
can be found in more detail in this volume~\cite{SarangiTalk}.
\subsection{The 0-lepton and 1-lepton modes}
\begin{figure}[h]
\centering
\includegraphics[width=80mm]{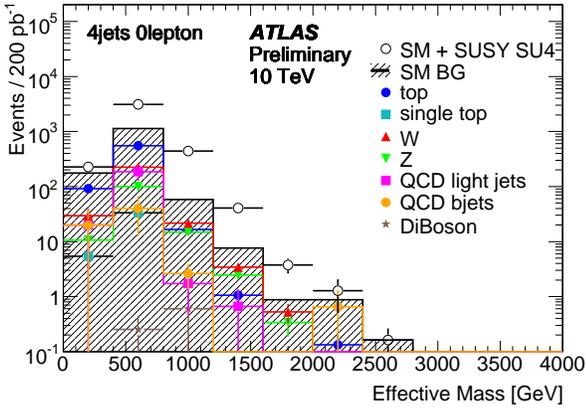}
\caption{\met distribution in the 0-lepton mode SUSY search for an integrated luminosity 
of 200~pb$^{-1}$ at 10 TeV. The open circles shows the signal (ATLAS mSUGRA benchmark point 
SU4) plus the total SM background while the different SM backgrounds are indicated in the legend.} 
\label{zerolepton}
\end{figure}

The 0-lepton mode (jets+$E_{\rm T}^{\rm miss}$) provides the least model dependent search. 
Events with at least four jets or three jets or two jets can be chosen. The main background 
in this mode is the QCD multijet background (\met is produced either by fluctuations in jet energy 
measurement or by $B$ hadron decays to real neutrino). A strong event selection cuts (requirement 
of high multiplicity jets, high \pt cuts on jets and a cut on the azimuthal angle between each 
of the first three jets and the $E_{\rm T}^{\rm miss}$) reduces this background significantly. The remaining 
background comes from $t \bar t$+jets, $W$+jets, $Z$+jets and diboson production. The global event 
variable, $M_{\rm eff}$, is used in discriminating SUSY from the SM events. \meff is defined as the scalar 
sum of the \pt of the leading jets and the \met and the \pt of the leptons (if present) considered 
in the analysis. \meff is also used to quantify the SUSY mass scale. From an ATLAS 
study~\cite{ATLASnoteInclusive10TeV}
the \meff distribution of the signal (ATLAS mSUGRA benchmark point, SU4, with parameters 
$m_0 =200$ GeV, $m _{1/2}=160$ GeV, $A_0=-400$, ${\rm tan}(\beta)=10$, ${\rm sgn}(\mu)=+$) 
and the total background are shown in Figure~\ref{zerolepton} for an integrated 
luminosity of 200~pb$^{-1}$ at 10 TeV. An excess of events is clearly observed in this mode.

\begin{figure}[h]
\centering
\includegraphics[width=80mm]{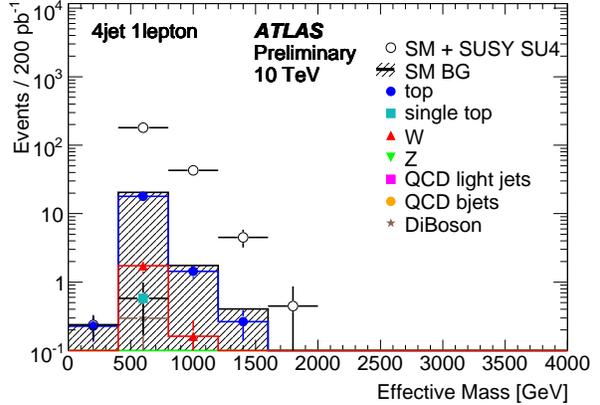}
\caption{\met distribution in the 1-lepton mode SUSY search for the same 
ATLAS study as in Figure ~\ref{zerolepton}.} 
\label{onelepton}
\end{figure}

The 1-lepton mode (jets+one lepton+$E_{\rm T}^{\rm miss}$) is often considered to be the golden mode for 
early SUSY searches. Requirement of the additional lepton (electron or muon) strongly 
reduces the QCD multijet background. The event selection cuts are similiar to that of the 
zero-lepton mode but includes a cut on the transverse mass of the lepton and the \met 
to suppress the $t \bar t$+jets and $W$+jets backgrounds. Figure~\ref{onelepton} shows 
the \met distribution for the same ATLAS study as in the 1-lepton mode. Although the 
statistics is reduced, the background is dominated by $t \bar t$+jets and $W$+jets which 
are expected to be better understood than the QCD background. Thus 1-lepton mode is 
more robust against uncertainities in the background estimations. 
\subsection{The 2-lepton and 3-lepton modes}
\begin{figure}[h]
\centering
\includegraphics[width=85mm]{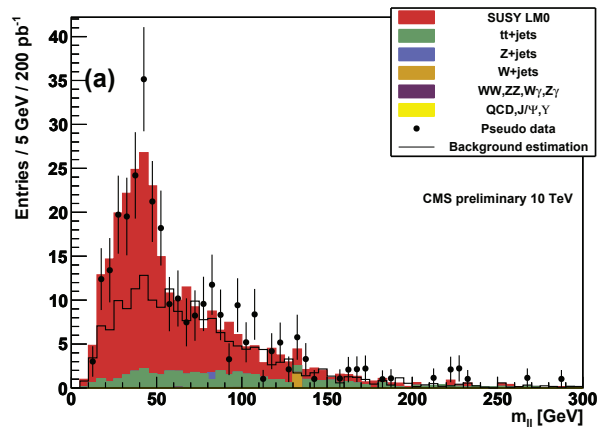}
\caption{The invariant mass of OSSF dileptons for an integrated luminosity of 
200~pb$^{-1}$ at 10 TeV. The red histogram shows the SUSY signal while the different 
SM backgrounds are indicated in the legend. The black solid line respresent the 
background from the OSOF pairs. The black dots shows the distribution when 
exactly 200~pb$^{-1}$ of Monte Carlo events are used, namely without scaling down 
from a larger sample.} 
\label{dilepton}
\end{figure}

The 2-lepton mode with jets+two leptons (opposite-sign or same-sign)+\met in the final state 
can have high discovery potential in early SUSY searches. The opposite-sign ($\mathrm {OS}$) 
dileptons arising from the SUSY decays 
(${\widetilde \chi}_{2}^{0} \rightarrow {\widetilde \chi}_{1}^{0}l^{\mp} l^{\pm}$) 
have the same-flavor to avoid $\mu \rightarrow e \gamma$ decay. The dileptons coming from 
different decay chains (SUSY combinatorial and the SM backgrounds) can be of the same-flavor 
(SF) or opposite-flavor (OF) with the same probability. Thus observing an excess of OSSF 
dilepton events over OSOF events would indicate a clear evidence for new physics. 
Figure~\ref{dilepton} shows the invariant mass of OSSF dileptons from a CMS 
study~\cite{CMSnoteDilepton10TeV} 
for an integrated luminosity of 200~pb$^{-1}$ at 10 TeV. The CMS mSUGRA benchmark point, 
LMO ($m_0 =200$ GeV, $m _{1/2}=160$ GeV, $A_0=-400$, ${\rm tan}(\beta)=10$, 
${\rm sgn}(\mu)=+$) is considered. The red histogram shows the contribution from SUSY events. 
The dominant background is found to be $t \bar t$+jets. A significant excess of OSSF dileptons 
events over OSOF events (shown by the black solid line) is seen with the first 
200~pb$^{-1}$ of the LHC data. 
The same-sign dileptons could provide an almost background free channel for SUSY searches. 
Events with two prompt, isolated and same-sign dileptons are rare in the SM while in SUSY 
the gluino is a Majorana particle so events with same-sign dileptons can be common. Though 
the event rates will be very small (from 10 to 100 events/fb$^{-1}$) a recent ATLAS 
study~\cite{ATLASnoteSS10TeV} discusses that gluinos and squarks with masses up to 500~GeV 
can be discovered with a $3\sigma$ signal significance for an integrated luminosity of 
200~pb$^{-1}$ at 10 TeV. 

The 3-lepton mode (three leptons+$E_{\rm T}^{\rm miss}$) to search for signals from direct gaugino 
(such as ${\widetilde \chi}_{2}^{0}$ and ${\widetilde \chi}_{1}^{+}$) pair production has 
perhaps been a best search mode at the Tevatron~\cite{Tevatron3lepton}. The trilepton signal 
can come from leptonic decay of pairs of heavy gauginos (produced directly or in the decay of heavier 
partner particles) through real or virtual $W^{\pm}$, $Z^{0}$ or ${\widetilde l}$ to 
leptons and a pair of LSPs. A general approach has been to consider two analyses; first with 
the selection of the 3-leptons+jet events (no cut on $E_{\rm T}^{\rm miss}$) and second with the 
selection of the 3-leptons+\met events. In an ATLAS study~\cite{Aad:2009wy}, the 3-leptons+jet 
analysis is found to be more sensitive than the 3-leptons + \met one for the mSUGRA 
benchmark points studied for an integrated luminosity of 1~fb$^{-1}$ at 14 TeV. 
\subsection{The tau- and $b$-jet modes}
As for the tau- and $b$-jet modes, they can be dominant in SUSY models with large 
${\rm tan}(\beta)$, thus they can help determining the underlying model parameters. 
The requirement of a reconstructed $\tau$ in the final state eliminates the QCD background 
and the remaining background is dominated by $t \bar t$+jets and $W$+jets. Though the lower 
efficiency and purity in $\tau$ reconstruction, $\tau$ decays provide complimentary info 
on the nature of new physics; ${\widetilde \tau}$ LSP, $e / \mu / \tau$ universality. 

\begin{figure}[h]
\centering
\includegraphics[width=85mm]{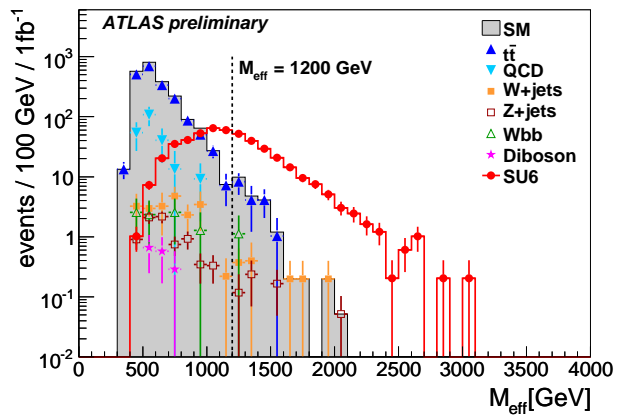}
\caption{The \meff distribution of the b-jet mode (two b-jets analysis) SUSY search for 
an integrated luminosity of 1~fb$^{-1}$ at 14 TeV. The signal point is the ATLAS mSUGRA 
benchmark point SU6. The contributions from the signal and the SM backgrounds are 
indicated in the legend.} 
\label{bjet}
\end{figure}

In mSUGRA models with large ${\rm tan}(\beta)$ the ${\widetilde b}$ and ${\widetilde t}$ 
are lighter than the first and second generation squarks. In this case ${\widetilde b}$ 
production is enhanced leading to decays with $b$-jets in the final state. A similiar 
analysis is performed as in the 0-lepton mode with the additional selection of at least 
two or three $b$-jets. The $b$-tagging performance at high \pt is an important 
issue for identifying $b$-jets. A recent ATLAS study~\cite{ATLASnotebjet} uses a 
$b$-tagging efficiency of $\sim$60$\%$ and a rejection of $\sim$100 ($\sim$10) 
against light quark ($c$-quark) jets and shows that requiring $b$-jets reduces 
the QCD multijets, $W$+jets and $Z$+jets backgrounds leaving the almost only background 
from $t \bar t$+jets. Figure~\ref{bjet} shows the \meff distribution of the mSUGRA 
signal point SU6 ($m_0 =320$ GeV, $m _{1/2}=375$ GeV, $A_0=0$, ${\rm tan}(\beta)=50$,
${\rm sgn}(\mu)=+$) together with the SM backgrounds for an integrated luminosity of 
1~fb$^{-1}$ at 14 TeV. A clear excess is seen for the signal events above the SM backgrounds.
\section{An Exclusive Search without \met}
A new approach to SUSY searches with dijet events has recently been proposed~\cite{Randall}. 
It focuses on a SUSY parameter space where squarks are pair produced and both decay 
directly to a quark and the lightest neutralino, leading to dijet events with \met in the final state. 
It exploits powerful discriminating variables to separate signal and background 
without making explicit use of the \met measurement, thus providing a very promising channel 
for early SUSY searches. A CMS study~\cite{CMSnoteMultijet} based on this approach is summarized 
in this section. The study has been carried out in the context of the mSUGRA model for an 
integrated luminosity of 100~pb$^{-1}$ at 10 TeV. 

\begin{figure}[h]
\centering
\includegraphics[width=80mm, height=60mm]{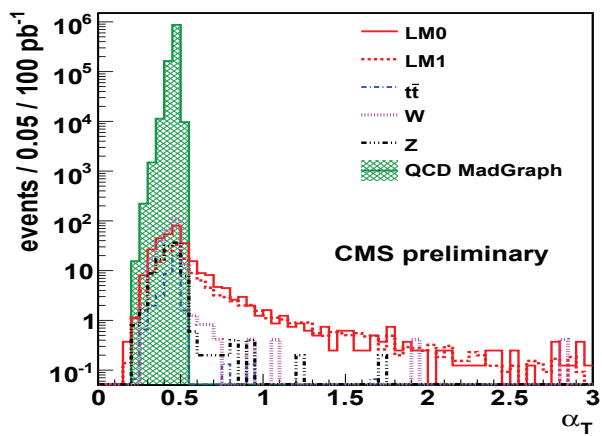}
\caption{The $\alpha_{\rm T}$ distribution for dijet events for an integrated luminosity
of 100~pb$^{-1}$ at 10 TeV. The labels LMO and LM1 represent the CMS mSUGRA benchmark 
points as their parameters given in the text. The SM backgrounds are also indicated in 
the legend.} 
\label{alphaT}
\end{figure}

The main backgrounds to the event topology are the QCD dijet events (\met is introduced through 
jet energy mismeasurements) and $Z$+jets events with $Z$ decaying into two neutrinos. A kinematic 
variable, $\alpha_{\rm T} = E_{\rm T}^{\rm j2} / M_{\rm T} $, is defined as the ratio of the transverse 
energy of the second jet and the transverse mass of the two jets. It exploits the requirement of 
back-to-back jets of equal magnitude for QCD events, thus $\alpha_{\rm T}$ is exactly 0.5 for 
well measured QCD dijet events. Figure~\ref{alphaT} shows the $\alpha_{\rm T}$ distribution for 
dijet events for signal and background. The mSUGRA benchmark points are labelled as LMO 
($m_0 =200$ GeV, $m _{1/2}=160$ GeV, $A_0=-400$, ${\rm tan}(\beta)=10$, ${\rm sgn}(\mu)=+$) 
and LM1 
($m_0 =60$ GeV, $m _{1/2}=250$ GeV, $A_0=0$, ${\rm tan}(\beta)=10$, ${\rm sgn}(\mu)=+$). 
Events are selected with $\alpha_{\rm T} > 0.55$ cut which removes 
almost all QCD background. The remaining small background is from $t \bar t$+jets, $W$+jets and 
$Z \rightarrow \nu \bar{\nu}$+jets events. This study also extends the dijet analysis to 
higher jet multiplicities where $n$ jets ($n=2...6$) are considered. It is found that the 
discriminating power of $\alpha_{\rm T}$ against QCD background provides signal over background 
ratios of 4 to 8 for favorable SUSY benchmark points. The results are shown to be robust against 
systematic uncertainties from the jet energy scale and jet direction. A data-driven approach 
has been taken to estimate both the QCD background and the remaining backgrounds. Even though 
the above Monte Carlo study suggests that the QCD background is negligible (after the 
$\alpha_{\rm T}$ cut), it has been verified by the data-driven estimates that it is indeed 
the case. The details of this study can be found in this volume~\cite{LunguTalk}. 
\section{Special SUSY Signatures}
A number of SUSY models predict specific signatures that may not be observed by the above 
traditional SUSY searches, such as long-lived massive particles or high \pt photons. For 
these specific signatures the SM backgrounds are often small thus their detection could provide 
unique constraints on the SUSY breaking mechanisms. A list of SUSY models giving rise to stable 
massive particles can be found in~\cite{MassiveStableParticles}. Examples are MSSM, GMSB, AMSB 
and Split-SUSY models. 

These particles live long enough to pass through the detector or decay in it. Such a particle 
can be a slepton (${\widetilde \tau}_{1}$, ${\widetilde e}_{R}$, ${\widetilde \mu}_{R}$ 
in GMSB model), a gluino (in Split-SUSY model), a stop (gravitino LSP scenario of SUGRA 
models~\cite{gravitinoLSPscenario}) or a lightest neutralino (in GMSB model). Their signatures will be 
different at the detector. If it is a stable charged slepton, its interaction 
with the detector will only be ionizations, its observed track will look like a muon, except 
for its higher energy deposition and longer time of flight than a muon.  If it is a gluino 
or a stop, it is meta-stable, it will form bound states, so-called R-hadrons. Their signatures 
will be similiar to that of the stable slepton together with the appearance of high \pt tracks 
in the muon system with no matching track in the inner detector or the electric charge 
flipping between the inner detector and the muon system. If it is a long lived neutralino, 
it can travel a significant distance before decaying into a photon and a gravitino. Such 
photons will not point back to the beam interaction point called as non-pointing photons. 
If the neutralino's lifetime is not too long, events with two high \pt photons (prompt photons, 
originate close to the beam interaction point) plus large \met (from gravitinos) are expected.

As an example one study on the detection of the long lived gluinos is summarized in what follows. 
The Split-SUSY model suggests a large mass splitting between the new scalars and new fermions, 
then gluinos can only decay through a highly virtual squark if R-parity is conserved. Gluinos 
can thus be long-lived and may well be stable on the typical experimental timescales. 
When produced these gluinos will hadronize into bound states (R-hadrons), ${\tilde g}g$, 
${\tilde g}q \bar q$, ${\tilde g}qqq$. A recent CMS study~\cite{CMSnoteGluino} discusses 
a search strategy to detect the decays of such R-hadrons. The charged R-hadrons will lose energy via 
ionization as they traverse the detector, this energy loss may be sufficient 
to stop a significant fraction of the particles inside the detector volume. These stopped 
R-hadrons may decay at times when there are no collisions (beam gaps) or no beam 
(interfill period) in the LHC. Observing such decays would immediately indicate a new 
physics signal.

\begin{figure}[h]
\centering
\includegraphics[width=80mm]{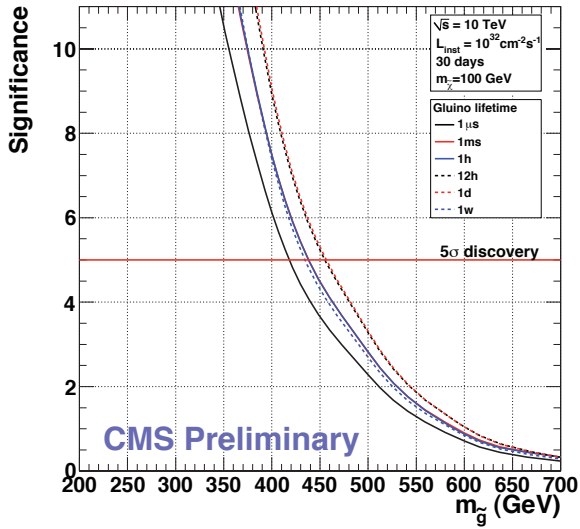}
\caption{The signal significance achievable after 30 days of running with
an instantaneous luminosity of $10^{31}$ cm$^{-2}$ at 10 TeV as a function of the
gluino mass from a counting experiment. The gluino lifetimes 
(from one micro second to one week) are indicated in the legend.} 
\label{longlivedgluino}
\end{figure}

A custom Monte Carlo simulation has been developed to estimate the signal efficiency 
and a novel triggering strategy has been implemented to improve the signal sensitivity. 
The only physics background is from cosmic rays. There is also an instrumental 
background. The uncertainties on the background determination are statistical and 
they are found to be small. However, the search sensitivity to a particulat model 
(Split-SUSY) involves some significant systematic uncertainties; theoretical uncertainties 
in the NLO calculation of the gluino-gluino production cross section, uncertainty on 
the stopping efficiency from the Monte Carlo simulation, uncertainty on modeling the 
exponential decay of the instantaneous luminosity. Figure~\ref{longlivedgluino} shows 
the signal significance that can be achieved after 30 days of running with 
an instantaneous luminosity of $10^{31}$ cm$^{-2}$ at 10 TeV as a function of the 
gluino mass from a counting experiment. A $5\sigma$ discovery should be possible 
in a matter of days for the models with large cross sections ($\sim$1 nb).  
\section{SUSY Discovery Reach}
In the previous sections examples of SUSY search strategies have been given 
based on the analyses of SUSY benchmark points with R-parity conservation. 
These points can not be representative of all possible SUSY models 
to be found at the LHC. They have been chosen with the aim of studying a 
variety of signatures and developing analysis techniques that can be applied 
to much of the SUSY parameters space accessible with early LHC data. In order 
to map the $5\sigma$ discovery reach of a SUSY model, a scan over its parameters 
is performed. 

\begin{figure}[h]
\centering
\includegraphics[width=85mm]{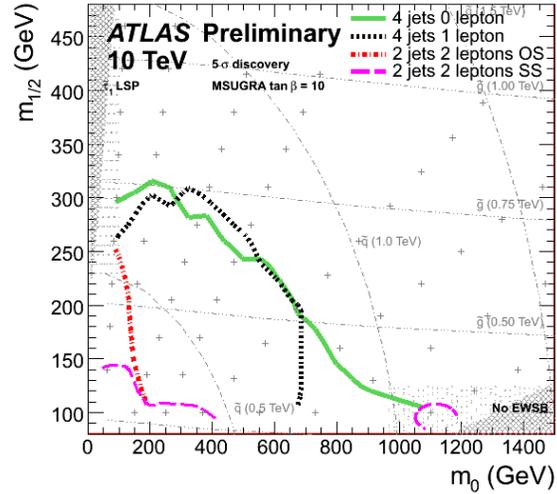}
\caption{The $5\sigma$ reach contours for the 0-lepton, 1-lepton
and 2-lepton search modes for mSUGRA as a function of $m_{0}$ and $m _{1/2}$.
The horizontal and curved grey lines indicate gluino and squark mass
contours, respectively. The scanned points are marked with a dagger. An 
integrated luminosity of 200~pb$^{-1}$ is assumed at 10 TeV.}
\label{mSUGRAscan}
\end{figure}

A recent ATLAS study~\cite{ATLASnoteInclusive10TeV} focuses on scanning the parameter space 
of mSUGRA, pMSSM (phenomenological MSSM with 19 free soft SUSY breaking parameters)~\cite{pMSSM} 
and UED (Universial Extra Dimensions)~\cite{UED} models considering an integrated luminosity 
of 200~pb$^{-1}$ at 10 TeV. Figure~\ref{mSUGRAscan} shows the $5\sigma$ discovery 
reach as a function of $m_{0}$ and $m _{1/2}$ parameters from a scan in mSUGRA 
with other parameters set to ${\rm tan}(\beta)=10$, $A_{0}=0$ and ${\rm sgn}(\mu)=+$. 
The signal Monte Carlo samples were produced using ATLFAST2 simulation 
(GEANT4 simulation of the inner detector, the muon system and fast simulation of the 
calorimeter). The top pair, single top and diboson samples were produced using GEANT4 
simulation while all other background samples with ATLFAST2. The reach plot has been 
made by finding the optimal \meff cut to maximize the signal significance in the 
0-lepton, 1-lepton and 2-lepton search modes.
This plot includes uncertainties on the SM background determination which are 
re-estimated from the data-driven methods studied in an earlier work~\cite{Aad:2009wy} 
for an integrated luminosity of 1~fb$^{-1}$ at 14 TeV. It is taken as $50\%$ for the 
overall background uncertainty in this study. As seen in the figure gluinos and 
squarks can be discovered with masses up to 750~GeV. The 0-lepton mode is found 
to be the best mode, however q-lepton mode is more robust against the QCD background. 
The reaches for the 2-lepton modes are less than the reaches for the 0-lepton and 
1-lepton modes. 

As for the $5\sigma$ discovery potential of the tau-mode, it is found to be slightly 
worse than the 0-lepton and 1-lepton modes due to the lower efficiency and purity 
in $\tau$ reconstruction which is studied in~\cite{Aad:2009wy} for 1~fb$^{-1}$ at 14 TeV.

The $5\sigma$ discovery potential of the $b$-jet mode is found to be comparable to that 
of the 1-lepton mode from a recent ATLAS study ~\cite{ATLASnotebjet} for 1~fb$^{-1}$ at 14 TeV. 
It is a competitive search mode specially at higher $m_{0}$ values. 

\begin{figure}[h]
\centering
\includegraphics[width=75mm]{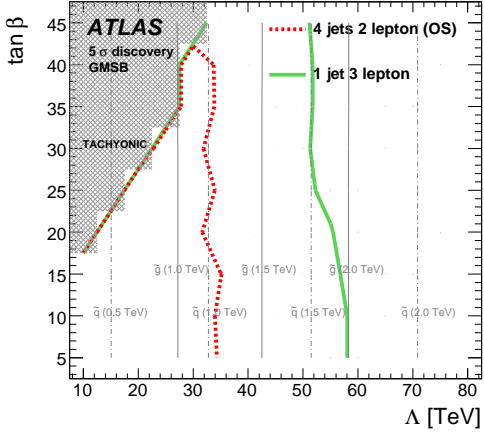}
\caption{The $5\sigma$ reach contours of the 2-lepton and 3-lepton 
analyses for the GMSB SUSY parameter scan as a function of ${\rm tan}(\beta$ 
and $\Lambda$. The vertical solid and dashed grey lines indicate gluino and 
squark masses respectively. An integrated luminosity of 1~fb$^{-1}$ is assumed 
at 14 TeV.} 
\label{GMSBscan}
\end{figure}

Scans in the NUHM (Non Universal Higgs Model)~\cite{NUHM} and GMSB SUSY parameter space have also 
been performed in an ATLAS study~\cite{Aad:2009wy} for 1~fb$^{-1}$ at 14 TeV. The discovery 
reach of the NUHM model is found to be vitually identical to that of mSUGRA in the 
0-lepton and 1-lepton modes. The $5\sigma$ discovery reach of the GMSB model as a function 
of ${\rm tan}(\beta)$ and $\Lambda$ parameters can be seen in Figure~\ref{GMSBscan} 
with other parameters set to $M_m$=500~TeV, $N_5$=5, $C_G$=1 and ${\rm sgn}(\mu)=+$. 
The reach for the 3-lepton analysis is significantly better than that of the 2-lepton analysis.
\section{SUSY Mass Measurements}
The searches strategies to establish SUSY discovery have been outlined in the 
previous sections. As soon as a SUSY discovery can be made by these searches, 
the emphasis will move on to confirming that it is really SUSY. The masses and 
spins of the newly discovered particles should be measured and the underlying
model parameters should be determined. Since it is not possible to cover all of 
the allowed SUSY models, mSUGRA model is taken as an example to develop the mass 
measurement techniques that can be applicable to large variety of SUSY and 
SUSY-like models. The study presented here is based on an ATLAS study 
~\cite{Aad:2009wy} performed for integrated luminosity of 1~fb$^{-1}$ at 14 TeV.

The mass measurement strategy is to exploit kinematics of the long decay chains. 
In the R-parity conserving mSUGRA models SUSY particles are pair produced and 
their cascade decays typically have high \pt jets, leptons and large \met (due to 
two invisible LSPs in every SUSY event) in the final states. No mass peaks can be 
reconstructed, however kinematic endpoints in the invariant mass distributions of 
dilepton, dijet, leptons+jets and ditau can be measured. These measurements are then 
used in deriving relations between SUSY masses. The theoretical positions of the 
endpoints are well known from the analytical expressions~\cite{Lester}. If sufficiently 
long decay chains can be isolated and enough endpoints can be measured
then the masses of the SUSY particles involved in the decay can be obtained in a 
model-independent way. As an example from the decay chain:
\begin{equation}
{\widetilde q}_{L} \rightarrow q {\widetilde \chi}_{2}^{0} \rightarrow q
{\widetilde l}_{R}^{\pm}l^{\mp} \rightarrow q {\widetilde \chi}_{1}^{0}
l^{\pm} l^{\mp}
\label{qLchain}
\end{equation}
the masses of ${\widetilde q}_{L}$, ${\widetilde l}_{R}$,
${\widetilde \chi}_{2}^{0}$ and ${\widetilde \chi}_{1}^{0}$ can be measured. This 
is the first decay chain likely to be reconstructed as it has the advantage of having 
charged leptons, \met and hadronic jets in the final state which ensures a large 
signal to background ratio. Also, the technique known as flavor subtraction allows to 
determine both the SUSY combinatorial and the SM background from the data itself with 
high accuracy. This technique exploits the fact that the signal contains two 
opposite-sign same-flavor (OSSF) leptons, while the background is due to pair of 
leptons coming from different decay chains, which can be of the same-flavor (SF) or 
opposite-flavor (OF) with the same probability. Thus the background cancels in 
the substraction 
$N({e}^{+}{e}^{-})/\beta + \beta N({\mu}^{+}{\mu}^{-}) -  N(e^\pm \mu^\mp)$
where $N$ indicates the respective number of events and $\beta$ is the ratio of
the electron and muon reconstruction efficiencies ($\beta$=0.86 for this study).
\subsection{Kinematic endpoints}
\begin{figure}
\centering
\includegraphics[width=80mm]{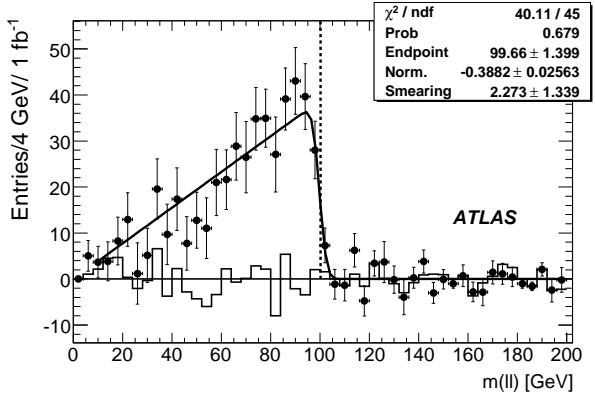}
\caption{The dilepton invariant mass distribution for the ATLAS benchmark point SU3 
with an integrated luminosity of 1~fb$^{-1}$ at 14 TeV. Different distributions and 
lines are explained in the text.}
\label{dilep}
\end{figure}

Starting with the two leptons (electron or muon) in the final state, 
the invariant mass distribution of dilepton exhibits a triangular shape with an endpoint at:
\begin{equation}
\label{endpointsu3}
\mlledge = m_{\chitwo}
      \sqrt{1-\left( \frac{m_{\slepton}}{m_{\chitwo}} \right)^2}
      \sqrt{1-\left( \frac{m_{\chione}}{m_{\slepton}} \right)^2 \ }
\end{equation}
In Figure~\ref{dilep}, the dilepton invariant mass distribution is shown 
after flavor subtraction applied for the ATLAS benchmark point SU3. 
The line histogram shows the small SM background contribution (dominated 
by $t \bar t$+jets), while the points are the sum of SM and SUSY contributions. 
A clear excess is seen together with a clear edge structure. The distribution is 
fitted with a triangle smeared with a Gaussian which is superimposed 
in the plot. Also, the expected (truth) position of the endpoint is indicated 
by a vertical dashed line. The fitted endpoint is $(99.7 \pm 1.4 \pm 0.3)$~GeV 
where the first error is statistical and the second is the systematic error on 
the lepton energy scale and the $\beta$ parameter (10\%). Compared to its truth 
value of 100.2~GeV (calculated from Eq.~\ref{endpointsu3}) the endpoint can be measured 
with a precision of a few percent already with 1~fb$^{-1}$ of integrated 
luminosity at 14 TeV for the model chosen.

In order to determine the masses of all the SUSY particles involved in the decay chain, 
further kinematic endpoints need to be measured from the mass distributions involving 
a jet: $\mllqedge$, $\mllqthr$, $\mlqlowedge$ and $\mlqhighedge$. The label thr indicates 
lower endpoint of the distribution (a non-zero threshold value) whereas low/high indicate 
minimum/maximum of the two masses $m_{l^{+}q}$ and $m_{l^{-}q}$. Only the two leading jets are 
considered in event selection as it is not possible to identify the quark from the 
${\widetilde q}_{L}$ decay. Therefore it is assumed that it hadronizes into one of the 
two highest $p_{T}$ jets in the event. The fitted endpoints together with their truth 
values can be seen in Table~\ref{fitresult}. 
The first error is statistical, the second and third are the lepton energy scale uncertainty 
and the jet energy scale uncertainty, respectively. While the measured 
values are compatible with their truth values, there are large uncertainties on the measured values. 
This is due to not having clear edge structure (combinatorics from choosing the wrong jet in a 
true SUSY event) and thus trying to fit the tails (beyond the endpoint) with straight lines.
\begin{table}[h]
\begin{center}
\caption{Measured endpoint values for the ATLAS mSUGRA benchmark point SU3 
with an integrated luminosity of 1~fb$^{-1}$ at 14 TeV. The truth values are also listed.}
   \begin{tabular}{l|c|c}
   \hline
   Endpoint & SU3 Measured $\;$(GeV) & SU3 Truth $\;$(GeV) \\
   \hline
    $\mllqedge$    & $517 \pm 30 \pm 10 \pm 13$ & 501  \\
    $\mllqthr$     & $265 \pm 17 \pm 15 \pm  7$ & 249  \\
    $\mlqlowedge$   & $333 \pm 6  \pm  6 \pm  8$ & 325  \\
    $\mlqhighedge$  & $445 \pm 11 \pm 11 \pm 11$ & 418  \\
   \hline
   \end{tabular}
\label{fitresult}
\end{center}
\end{table}

In a similiar way, the endpoint of the invariant mass distribution of ditau, 
$m_{\tau \tau}^{\rm edge}$, can be measured from the decay 
$\chitwo \rightarrow {\widetilde \tau}_{1} \tau \rightarrow \chione \tau^{\pm} \tau^{\mp}$. 
The fitted endpoint is $(102 \pm 17^{\rm stat} \pm 5.5^{\rm syst} \pm 7^{\rm pol})$~GeV 
for the ATLAS benchmark point SU3. 
The first error is statistical, the second is the systematic uncertainty dominated by the fitting 
procedure and the third is the uncertainty from the polarization effects. Its truth value 
is 98.3~GeV. 
 
Also, from the decay ${\widetilde q}_{R} \rightarrow \chione q$ the mass of the right-handed 
squark can be measured from the endpoint of the stransverse mass distribution by assuming that 
the mass of $\chione$ is known from the dilepton and leptons+jets endpoint measurements. 
The fitted endpoint is $590\pm 9(\rm stat)^{+13}_{-6}(\rm syst)$ for the ATLAS benchmark point 
SU3. The systematic uncertainty accounts for the choice of the fit limits and the jet energy 
scale variations. It agrees well with its truth value of 611~GeV.
\subsection{Extraction of SUSY masses from endpoint measurements}
From a combination of experimentally measured endpoints as above the SUSY particle masses 
can be extracted. By using the five endpoint measurements, $\mlledge$, $\mllqedge$, 
$\mllqthr$, $\mlqlowedge$, $\mlqhighedge$, the four unknown SUSY 
particle masses involved in the decay, ${\widetilde q}_{L}$, ${\widetilde l}_{R}$, ${\widetilde \chi}_{2}^{0}$, 
${\widetilde \chi}_{1}^{0}$ can be solved. A numerical $\chi^2$ minimization based on the MINUIT 
package is performed. The masses resulting from the $\chi^2$ fit can be seen in Table~\ref{masses}
for the SU3 point. The first error is statistical, the second accounts for the jet energy
scale uncertainty. An anticorrelation with the jet energy scale variation is indicated 
by $\mp$ sign. The reconstructed masses are found to be highly correlated 
with $\chione$ which is not well determined by the endpoint techniques. 
Therefore the precision on the absolute mass values is rather moderate. However, 
the mass differences are better measured than the absolute masses as expected. 
\begin{table}[h]
\begin{center}
\caption{Reconstructed SUSY particle masses and mass differences 
together with their truth values for the SU3 benchmark point with 
1~fb$^{-1}$ at 14~TeV.}
\begin{tabular}{l|c|c}
\hline
Observable & SU3 Reconstructed $\;$(GeV)&  SU3 Truth $\;$(GeV)\\
\hline
$m_{\chione} $  &   $88 \pm 60 \mp 2$ & 118 \\
$m_{\chitwo} $  &  $189 \pm 60 \mp 2$ & 219 \\
$m_{\squark}  $  &  $614 \pm 91 \pm 11$& 634 \\
$m_{\slepton} $  &  $122 \pm 61 \mp 2$ & 155 \\
\hline
$m_{\chitwo} - m_{\chione}$ &  $100.6 \pm 1.9 \mp 0.0$  & 100.7 \\
$m_{\squark} - m_{\chione}$  &  $526  \pm 34 \pm 13$     & 516.0 \\
$m_{\slepton} - m_{\chione}$ &  $34.2 \pm 3.8 \mp$ 0.1   &  37.6 \\
\hline
\end{tabular}
\label{masses}
\end{center}
\end{table}
\section{Determination SUSY Model Parameters}
The ultimate goal is to pin down the SUSY model parameters.
The SUSY parameter-fitting package Fittino~\cite{Fittino} has been used to perform the
calculations of the model parameters from all the endpoints measured;
$\mlledge$, $\mllqedge$, $\mllqthr$, $\mlqlowedge$, $\mlqhighedge$, $m_{\tau \tau}^{\rm edge}$,
$m_{\rm T2}({\widetilde q}_{R})$.
The fitted mSUGRA parameters can be seen in Table~\ref{paramfits} for the SU3 point. Already 
with 1~fb$^{-1}$ of data $m_0$ and $m _{1/2}$ parameters can be derived reliably, however 
determination of ${\rm tan}(\beta)$ and $A_0$ is more problematic since no information from the 
Higgs sector is available at low luminosity. 
\begin{table}[h]
\begin{center}
\caption{Fitted mSUGRA parameters for the SU3 point with 1~fb$^{-1}$ at 14 TeV. ${\rm sgn}(\mu)=+$.}
    \begin{tabular}{l|c|c}
      \hline
      Parameter & SU3 Fitted  & SU3 Truth\\
      \hline
      ${\rm tan}(\beta)$ &  $7.4\pm  4.6$ & $6$ \\
      $m_{0}$ (GeV)      &  $98.5\pm 9.3$ & $100$ \\
      $m_{1/2}$ (GeV)     &  $317.7\pm 6.9$ & $300$\\
      $A_{0}$ (GeV)       &  $445\pm 408$ & $-300$   \\
      \hline
    \end{tabular}
\label{paramfits}
\end{center}
\end{table}
\section{Conclusions}
If SUSY particles exist in nature at sub-TeV mass range the ATLAS and CMS experiments 
will open up a discovery window for new physics beyond the Standard Model. Search 
strategies have been developed to cover a broad range of signatures expected from 
SUSY models. Understanding the detector performance and controlling the SM backgrounds 
via data-driven methods are important tasks for early SUSY analyses. Once a 
signature consistent with SUSY has been established then various measurements 
will have to be combined to reconstruct the SUSY mass spectrum and to understand 
the SUSY breaking mechanism. 

Early SUSY searches with 100/200~pb$^{-1}$ integrated luminosity at 10 TeV center of 
mass energy suggest that we are already sensitive to an unexplored area in less than 
a year running. Exciting times are ahead with the upcoming start-up of the LHC.

\end{document}